\documentclass[aps,prd,preprint,preprintnumbers,unsortedaddress,superscriptaddress,showpacs,nofootinbib]{revtex4-1}

\pdfoutput=1

\usepackage{graphicx}
\usepackage{amsmath}
\usepackage{amsfonts}
\usepackage{amssymb}
\usepackage{color}

\usepackage[colorlinks, citecolor=blue,anchorcolor=red,menucolor=red, linkcolor=red,filecolor=red,runcolor=red,urlcolor=blue,frenchlinks=red]{hyperref}

\begin{document}%\bibliographystyle{unsrt}
\arraycolsep1.5pt

\title{The $B^+ \to J/\psi \omega K^+ $ reaction and  $D^*\bar{D}^*$  molecular states}

\author{L.~R.~Dai}
\email{dailr@lnnu.edu.cn}
\affiliation{Department of Physics, Liaoning Normal University, Dalian 116029, China}
\affiliation{Departamento de F\'isica Te\'orica and IFIC, Centro Mixto Universidad de Valencia-CSIC,
Institutos de Investigac\'ion de Paterna, Aptdo. 22085, 46071 Valencia, Spain
}

\author{G. Y. Wang}
\affiliation{School of Physics and Engineering, Zhengzhou University, Zhengzhou, Henan 450001, China}

\author{X. Chen}
\affiliation{Department of Physics, Liaoning Normal University, Dalian 116029, China}

\author{E.~Wang}
\email{wangen@zzu.edu.cn}
\affiliation{School of Physics and Engineering, Zhengzhou University, Zhengzhou, Henan 450001, China}

\author{E.~Oset}
\email{oset@ific.uv.es}
\affiliation{Departamento de F\'isica Te\'orica and IFIC, Centro Mixto Universidad de Valencia-CSIC,
Institutos de Investigac\'ion de Paterna, Aptdo. 22085, 46071 Valencia, Spain
}

\author{D. M. Li}
\affiliation{School of Physics and Engineering, Zhengzhou University, Zhengzhou, Henan 450001, China}

\date{\today}
\begin{abstract}
We study the $B^+ \to J/\psi \omega K^+ $  reaction, and show that it is driven by the presence of two resonances,
the $X(3940)$ and $X(3930)$, that are of molecular nature and couple most strongly to $D^* \bar{D}^*$, but also to $J/\psi\omega$. Because of that, in the
$J/\psi\omega$ mass distribution we find a peak  related to the excitation of the resonances and a cusp with large strength at the $D^* \bar{D}^*$ threshold.
\end{abstract}

\maketitle

%\section{Introduction}
%\input{intro-ew.tex}
\section{Introduction}
\label{sec:intro}
In the last decades, many hadron resonances have been observed experimentally, and some of them can not be explained as conventional mesons and baryons\cite{PDG2018}. There exist different theoretical interpretations on the nature of these states, such as molecular, hybrid, multi-quark state, threshold enhancements, and so on~\cite{Chen:2016qju}. As one of the popular interpretations, molecular states have long been an important subject in hadron physics~\cite{Guo:2017jvc}.

However, it is not always easy to firmly identify some states as of molecular nature because of the existence of other interpretations, such as standard $q\bar{q}(qqq)$ or multi-quark states~\cite{Chen:2016qju,Karliner:2017qhf}.
One of the defining features associated to the molecular states that couple to several hadron-hadron channels is that one can find a strong and unexpected cusp in one of the weakly coupled channels at the threshold of the channels corresponding to the main component of the molecular state~\cite{Wang:2017mrt,Dai:2018tgo}.

A recent example of this feature is found in the $B^+\to J/\psi\phi K^+$ decay measured by the LHCb collaboration~\cite{Aaij:2016nsc,Aaij:2016iza}.
The LHCb analysis, including only the $X(4140)$ resonance at low $J/\psi\phi$ invariant masses, results in a width for the $X(4140)$ resonance much  larger than the average of the PDG~\cite{PDG2018}.  We have studied this reaction, taking into account the molecular state $X(4160)$, in addition to the $X(4140)$ resonance, and provided a better description of the low $J/\psi\phi$ mass distribution~\cite{Wang:2017mrt}.
According to Ref.~\cite{Molina:2009ct}, the $X(4160)$ state is a $D^*_s\bar{D}^*_s$ state with $I^G(J^{PC})=0^+(2^{++})$ and couples to $J/\psi\phi$. As a result, the $J/\psi\phi$ mass spectrum develops a strong cusp at the $D^*_s\bar{D}^*_s$ threshold. A similar structure is also seen, although with poor statistics, in the recent BESIII work on the $e^+e^-\to \gamma J/\psi \phi$~\cite{Ablikim:2017cbv}, and the corresponding discussion can be seen in Ref.~\cite{Wang:2018djr}.

In Ref.~\cite{Molina:2009ct}, two $D^*\bar{D}^*$ molecular states were found as $I^G(J^{PC})=0^+(0^{++})$ and $0^+(2^{++})$ at 3943 and 3922~MeV, respectively.
In addition to the strongly coupled channel $D^*\bar{D}^*$, both states also couple to $J/\psi\omega$ in the second place. In order to show the features of the molecular nature of these two resonances, we studied the $B^-_c \to \pi^- J/\psi\omega$ reaction, and found a peak around $3920 \sim 3940$~MeV, corresponding to the excitation of these two resonances, and a cusp with large strength at the $D^*\bar{D}^*$ threshold, in the $J/\psi\omega$ invariant mass distribution~\cite{Dai:2018tgo}.

In Ref.~\cite{Andreassi:2014skr} one finds  the $J/\psi \omega$ invariant mass distribution for the $B^+ \to J/\psi \omega K^+$ reaction 
using data collected by the LHCb experiment~\footnote{We are thankful to G. Andreassi for allowing us to use the data of his PhD dissertation~\cite{Andreassi:2014skr}.}, which is used to search for exotic states, in particular the $X(3872)$. If we look at Fig.~4.1 of Ref.~\cite{Andreassi:2014skr}, a peak presumably due to the excitation of $X(3872)$ is observed, but another peak around $3920\sim 3940$~MeV and a cusp-like structure around the 4020~MeV, the threshold of $D^*\bar{D}^*$, are also seen, which are similar to the features we found in the $J/\psi\omega$ invariant mass distribution of the $B^-_c \to \pi^- J/\psi\omega$ reaction~\cite{Dai:2018tgo}. It should be noted that, in Ref.~\cite{Andreassi:2014skr}, the peak around $3920\sim 3940$~MeV
is associated to the $\chi_{c0}(2P)$, with the fitted mass 3915~MeV. However, there is a debate about the mass of the $\chi_{c0}(2P)$~\cite{Chen:2016qju,Eichten:2004uh}, and the decay channel $J/\psi \omega$ of $\chi_{c0}(2P)$ is OZI suppressed.  In addition, prior to Ref.~\cite{Andreassi:2014skr}, Belle and Babar collaborations also found peaks in the $3920\sim3940$~MeV region of the $J/\psi \omega$ mass distribution for this reaction~\cite{Abe:2004zs,Aubert:2007vj}, and the resonance associated to the peak has a mass of $3943\pm11\pm13$~MeV in the analysis of Belle collaboration~\cite{Abe:2004zs}.~\footnote{Because of the large bin size of the Belle and Babar data ($\sim 50$~MeV), both the $J/\psi\omega$ mass distributions do not show a significant cusp  around the $D^*\bar{D}^*$ threshold, and we do not consider the data of Belle and Babar collaboration in this work.} Thus, it is straightforward to analyze the $B^+\to J/\psi\omega K^+$ reaction by considering the two $D^*\bar{D}^*$ molecular states found in Ref.~~\cite{Molina:2009ct}.

In this work, we will investigate the $J/\psi\omega$ interaction in the $B^+\to J/\psi\omega K^+$ reaction, and show that the peak around $3920\sim 3940$~MeV,
together with the cusp around $D^*\bar{D}^*$ threshold, can be tied to the  $D^*\bar{D}^*$ molecular structure of the two states.

This paper is organized as follows. In Sec.~\ref{sec:form}, we present the mechanisms of the $B^+\to J/\psi\omega K^+$ reaction, our results and discussions
are given in Sec.~\ref{sec:res}. Finally, a short summary is given in Sec.~\ref{sec:conc}.

\section{Formalism}
\label{sec:form}
\begin{figure}
\includegraphics[width=0.55\textwidth]{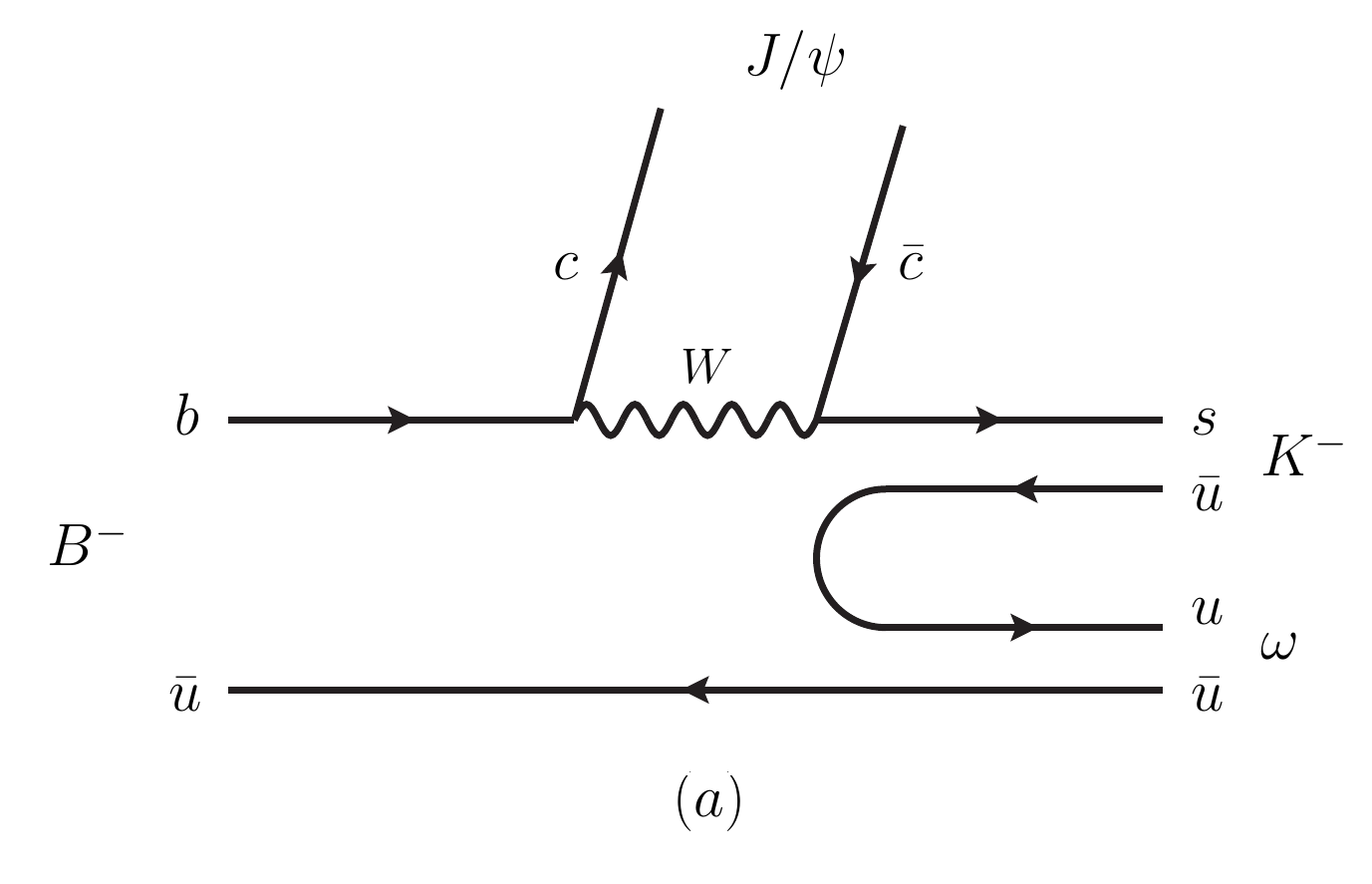} \includegraphics[width=0.55\textwidth]{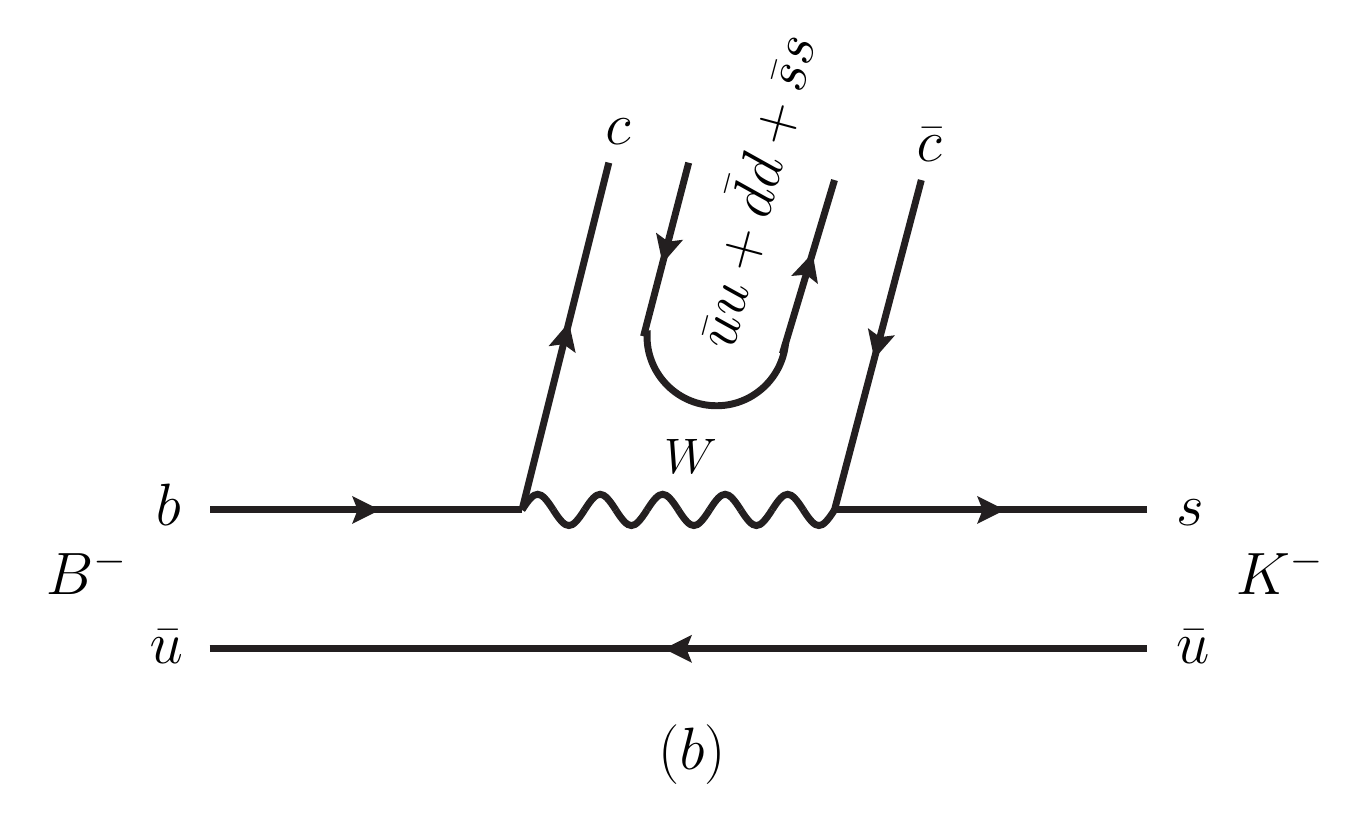} \includegraphics[width=0.55\textwidth]{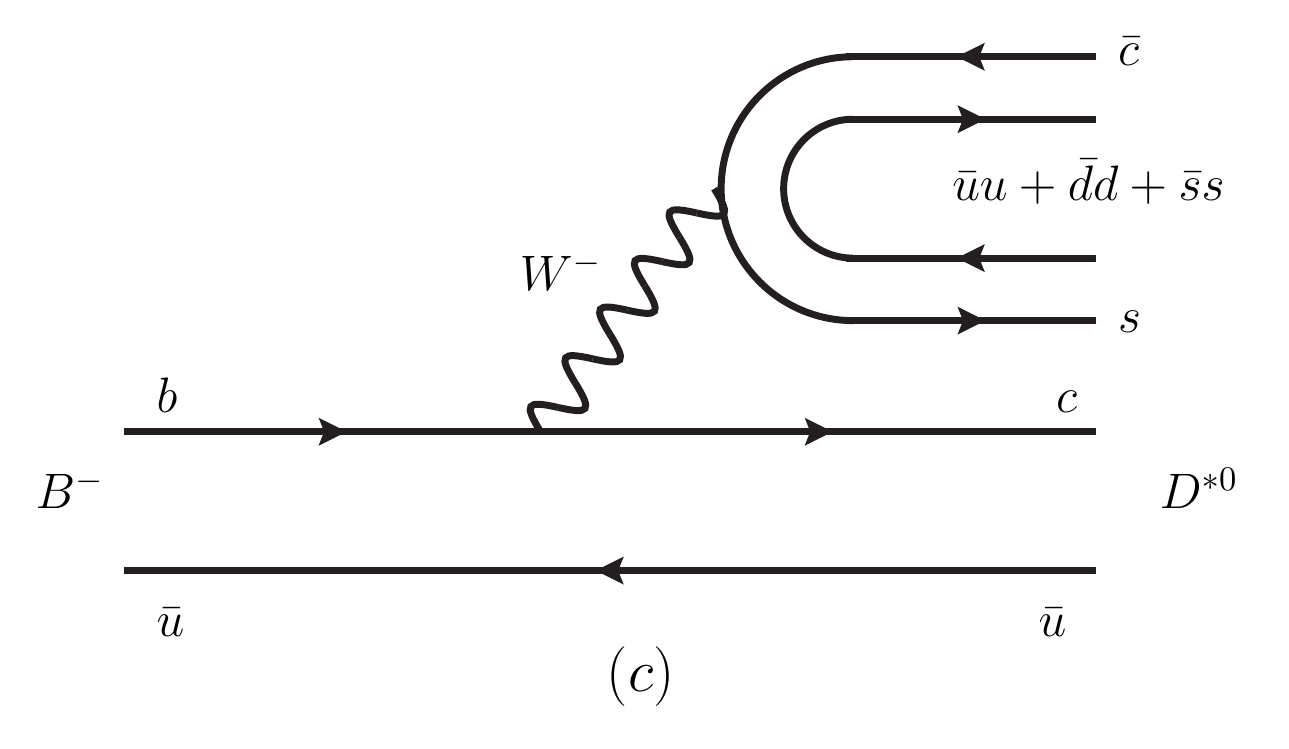}
\caption{\label{diag1} The internal emission mechanism at the  microscopic quark picture. (a) The $B^- \to J/\psi s\bar{u}$  decay and hadronization of $s\bar{u}$  through $\bar{q}q$
creation with vacuum quantum numbers; (b) $B^- \to K^- c\bar{c}$ decay  and  hadronization of $c\bar{c}$  through $\bar{q}q$; (c) The external emission mechanism and
 hadronization of the $s\bar{c}$ pair with $q\bar{q}$.}
\end{figure}

In order to deal with a quark $b$ rather than $\bar{b}$, we study the complex conjugate reaction of $B^+ \to J/\psi \omega K^+ $, $B^- \to J/\psi \omega K^-$.
 To see how the $B^- \to J/\psi \omega K^- $ reaction proceeds,  we look at the  possible Cabibbo-favored  mechanisms at the quark level.
 They  are depicted in Figs.\ref{diag1}(a), \ref{diag1}(b) and \ref{diag1}(c).
 The mechanisms of Figs.\ref{diag1}(a) and \ref{diag1}(b) imply internal  emission followed by hadronization of a primary $q \bar{q}$ component by an extra $\bar{q}q$ component with the
quantum numbers of  the vacuum.
In Fig.\ref{diag1}(a) the primary $s\bar{u}$ component  is hadronized  with $u\bar{u}$ and we then get $\omega K^-$ together with $c\bar{c}$  that leads to $J/\psi$.  In Fig.\ref{diag1}(b)  we  hadronize the $c\bar{c}$ component and we  can get $D^{*} \bar{D}^{*}$ together with $s\bar{u}$ that leads to $K^-$. This is not the $J/\psi \omega K^-$  state that one  observes, but the idea  is that  the $D^{*} \bar{D}^{*}$  will interact  and create a resonance  that couples to $J/\psi \omega$.

 In Fig.\ref{diag1}(c) we have the mechanism of external emission, which is color-favored with respect to internal emission. Here the primary $s\bar{c}$ pair is hadronized and we get
 also  a $\bar{D}^{*0} K^-$  component  together with $D^{*0}$ from the remaining $c\bar{u}$ pair.  Once again the $\bar{D}^{*0} D^{*0}$ pair interacts to produce $J/\psi \omega$ at the end.

\begin{figure}
\includegraphics[width=0.55\textwidth]{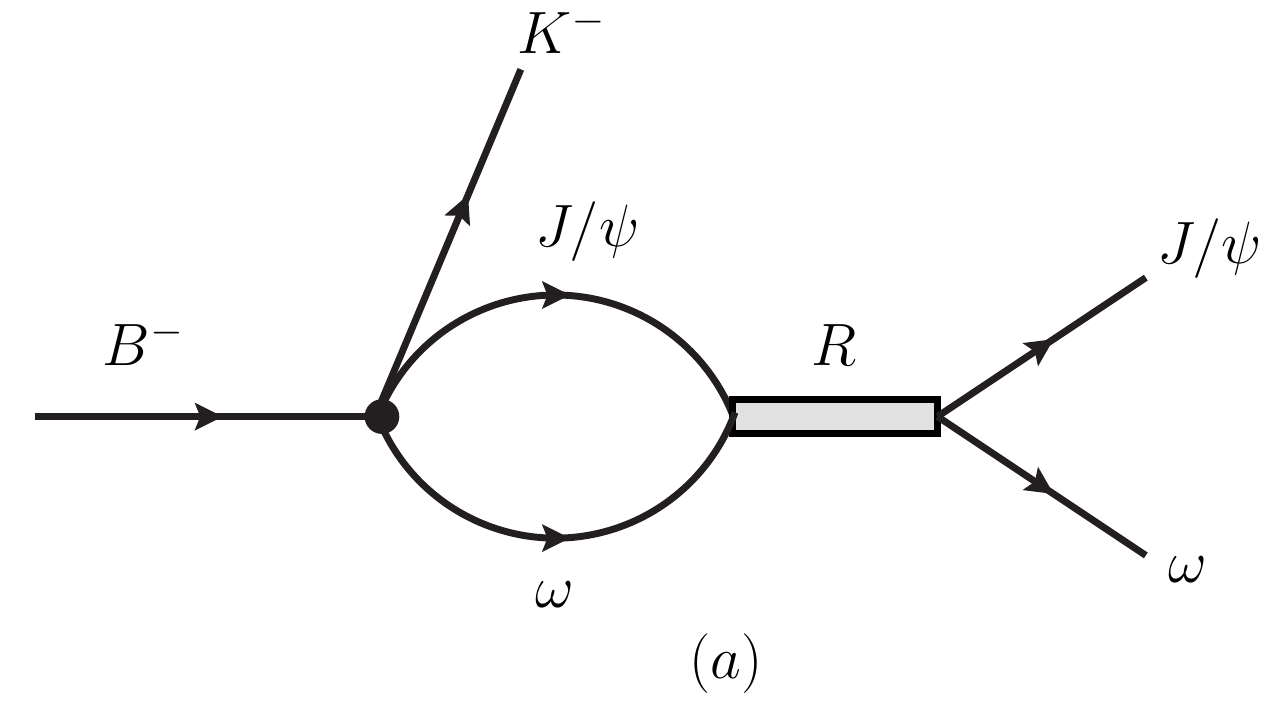} \includegraphics[width=0.55\textwidth]{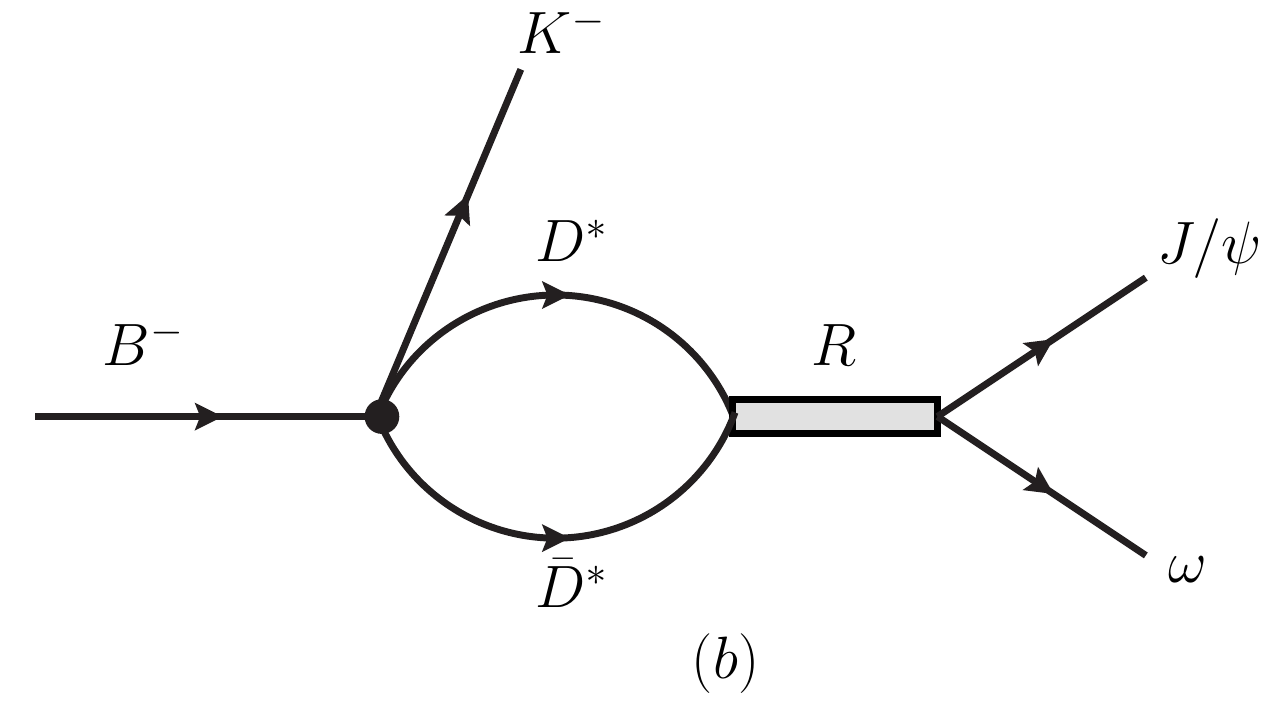}
\caption{\label{resc} Interaction mechanism to produce the $J/\psi\omega$ final state (a) through rescattering of  $J/\psi\omega$  components;
(b) through rescattering of $D^{*} \bar{D}^{*}$ components. $R$ is either the $X(3922)\,(2^{++})$ or $X(3943)\,(0^{++})$.}
\end{figure}

\begin{figure}
\includegraphics[width=0.6\textwidth]{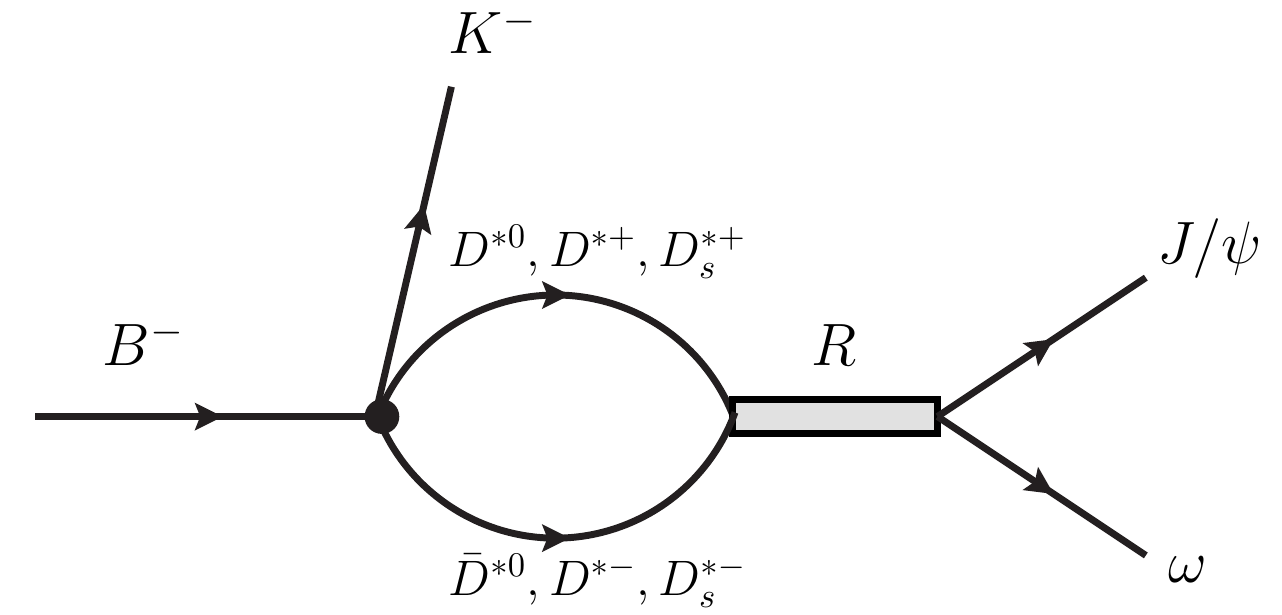}
\caption{\label{res2} Mechanism to produce the $J/\psi\omega$ final state
	through rescattering of $D^{0} \bar{D}^{0}$, $D^{*+} \bar{D}^{*-}$, and $D_s^{*+} \bar{D_s}^{*-}$  components.}
\end{figure}

 It is instructive  to see which are the most important mechanisms  of those discussed above. One is tempted  to choose the one in Fig.\ref{diag1}(a) that produces the $ J/\psi \omega K^- $
 without the need of rescattering. However,  since the interaction of $D^{*} \bar{D}^{*}$ produces resonances, the rescattering  mechanism becomes more important than the tree level  direct  production
 in the energy region of the $D^* \bar{D}^*$ resonant states.  To see  this, let us  see the meson decomposition of Fig.\ref{diag1}(b). Hadronizing with  $\bar{u} u$  the  $c\bar{c}$ component we get $D^{*0}\bar{D}^{*0}$, hadronizing with $\bar{d}d$ we obtain $D^{*+}{D}^{*-}$ and  hadronizing with $\bar{s}s$ we obtain $D_s^{*+}{D_s}^{*-}$, and, hence, we get the combination $D^{*0}\bar{D}^{*0} + D^{*+}{D}^{*-} + D_s^{*+}{D_s}^{*-}$. With the phase convention for isospin $(D^{*+},-D^{*0})$, $(\bar{D}^{*+},D^{*-})$ we have the isospin $I=0$ combination.
\begin{eqnarray}\label{eq:phase}
D^{*0}\bar{D}^{*0}(I=0) = \frac{1}{\sqrt{2}} \big(D^{*+} D^{*-} + D^{*0}\bar{D}^{*0}\big) \, .
\end{eqnarray}
Then  we look at the rescattering diagrams of   Fig.\ref{resc} creating the $I^G[J^{PC}]=0^+[0^{++}]$ resonance $X(3940)$  and decaying to $J/\psi \omega$.
With the  $J/\psi \omega$ and $D^{*} \bar{D}^{*}$  loop functions of the same size, the ratio of amplitudes of  Fig. \ref{resc} (a) versus  Fig.\ref{resc} (b) is
\begin{eqnarray}
\left|\frac{t(2a)}{t(2b)}\right| \simeq \left|\frac{g_{R,J/\psi \omega}}{\sqrt{2}g_{R,D^{*} \bar{D}^{*}}(I=0)}\right| \, .
\end{eqnarray}
In Table \ref{tab:g1} we give the couplings of this resonance to the different coupled channels  extracted from \cite{molina}. We find then that
\begin{eqnarray}\label{eq:ratio}
\left|\frac{t(2a)}{t(2b)}\right| \simeq 5 \% \, .
\end{eqnarray}
We see that this magnitude is small and as a consequence, we neglect the mechanism of  Fig.\ref{diag1}(a) and Fig.\ref{resc} (a) from our study.
A similar conclusion  can be reached  about the tensor  $0^+[2^{++}]$  resonance $X(3930)$. We  show the couplings of this latter resonance to the different coupled channels of  \cite{molina}
in Table \ref{tab:g2}.  We have neglected  the  $D_s^{*+}{D_s}^{*-}$ channel  for the counting of Eq.\eqref{eq:ratio}, which, if considered,  renders the ratio of Eq.\eqref{eq:ratio} even smaller.
The diagram of Fig.\ref{diag1}(c) corresponds to the color-favored external emission mechanism, which is usually enhanced  by a factor of three with respect to the
internal emission \cite{chau} and also usually interferes  constructively \cite{liang}. We will assume that  this is the case here too,  but will play  with uncertainties  on the relative weight
 of both mechanisms. With the hadronization coming from  Figs. \ref{diag1}(b) and \ref{diag1}(c) that we consider,  we have the hadronic state $|H\rangle$  produced before
 rescattering
\begin{eqnarray}\label{eq:h}
\big| H \big\rangle &=& \big| (D^{*0}\bar{D}^{* 0} + D^{*+}\bar{D}^{*-} + D_s^{*+}\bar{D}_s^{*-} + 3 \,C\, D^{*0}\bar{D}^{* 0}) K^- \big\rangle  \nonumber\,  \\
&=& \big|\left[(1+ 3 \,C)D^{*0}\bar{D}^{* 0} + D^{*+}\bar{D}^{*-} + D_s^{*+}\bar{D}_s^{*-} \right] K^- \big\rangle  \, .
\end{eqnarray}
With these primary components we shall take into account rescattering, which is depicted in Fig.~\ref{res2}, where $R$  will stand for any of the two resonances that we
produce with these channels. We will vary the value of $C$ around unity, but we can  anticipate that it hardly changes the shape of the distribution obtained. The minor
changes of the shape by changing  $C$ are due to the presence  of the $D_s^{*+}\bar{D}_s^{*-}$  channel,  which is not very important  in the present case.
If we remove this channel the results do not depend on $C$, except for a global normalization.
\begin{table}[h!]
\renewcommand\arraystretch{1.0}
 \begin{center}
 \caption{Couplings $g_{i}$ of the $0^{++}$  resonance to the relevant channels, in units of MeV.}
\label{tab:g1}
\centerline{$\sqrt{s}_{pole}=3943 + i 7.4$, $I^G[J^{PC}]=0^+[0^{++}]$}
\vspace{0.2cm}
\begin{tabular}{cc|cc}
\hline\hline
~~~channels~~~ &~~~ $g_i$~~~ & ~~~channels ~~~&~~~ $g_i$~~~ \\
\hline
$D^*\bar{D}^*$& $18810-i 682 $  & $\phi\phi$  & $-1000 -i 150$   \\
\hline
$D^*_s\bar{D}_s^*$ & $8426+i 1933 $ &$J/\psi J/\psi$ &$417+ i 64$ \\
\hline
$K^*\bar{K}^*$ & $10- i 11$  & $\omega J/\psi$ & $-1429 - i 216$  \\
\hline
$\rho\rho$ &  $-22 + i 47$   & $\phi J/\psi$ & $889+ i 196 $  \\
\hline
$\omega\omega$ & $1348+ i 234 $  & $\omega\phi$  &$-215 - i107$  \\
\hline\hline
\end{tabular}
\end{center}
\end{table}
\begin{table}[h!]
\renewcommand\arraystretch{1.0}
 \begin{center}
 \caption{Couplings $g_{i}$ of the $2^{++}$  resonance to the relevant channels, in units of MeV.}
\label{tab:g2}
\centerline{$\sqrt{s}_{pole}=3922+i 26$, $I^G[J^{PC}]=0^+[2^{++}]$}
\vspace{0.2cm}
\begin{tabular}{cc|cc}
\hline\hline
~~~channels~~~ &~~~ $g_i$~~~ & ~~~channels ~~~&~~~ $g_i$~~~ \\
\hline
$D^*\bar{D}^*$& $21100- i 1802$  & $\phi\phi$  & $-904 - i1783$  \\
\hline
$D^*_s\bar{D}_s^*$ & $1633+i 6797$ &$J/\psi J/\psi$ &$1783 +i 197$\\
\hline
$K^*\bar{K}^*$ & $42+ i 14 $  & $\omega J/\psi$ & $-2558 - i2289$  \\
\hline
$\rho\rho$ &  $-75 +i 37$  & $\phi J/\psi$ & $918+ i2921 $  \\
\hline
$\omega\omega$ & $1558 + i 1821$  & $\omega\phi$  & $91 -i 784$ \\
\hline\hline
\end{tabular}
\end{center}
\end{table}

%the results will not depend much on $c$ because the only
%difference will be the relative  of but this does not contribute to our two resonances.
%It does to $X(4160)$ as we  discussed in \cite{wang}.
%with $C=1,5,1,0.7$  to see the difference.  As in \cite{dai} we have to add the final state interaction

The combination of $| H \rangle$ in Eq.~\eqref{eq:h} accounts only for the flavor composition. We need
to take into account the spin-angular momentum structure of the vertices.
For  production of two vectors in $J=0$ we proceed in $L=0$ (for the kaon) to conserve angular momentum,  and we have the amplitude for the
$B^- \to  K^- D^{*+} \bar{D}^{*-}$

\begin{equation}\label{A}
A^{\prime}\, \vec{\epsilon} \cdot \vec{\epsilon}^{\,\,\prime}\, ,
\end{equation}
with $\vec{\epsilon}$, $\vec{\epsilon}^{\,\,\prime}$ the polarization vectors of $D^*$, $\bar{D}^*$. Note
that we shall work in the rest frame of the resonances produced, where $D^*$, $\bar{D}^*$ momenta are
small with respect to their masses and then we neglect the $\epsilon^0$ component.  This is actually very accurate for these momenta as can be seen in Appendix A of \cite{sakairamos}.

On the other hand, in the channel of $VV$ with $J=2$  we need $L=2$  to match the  angular momentum of the $B^-$,  and
this leads to the amplitude
\begin{equation}
B\, (\vec{\epsilon} \cdot \vec{k} \vec{\epsilon}^{\,\,\prime}\cdot \vec{k} -
\frac{1}{3} |\vec{k}|^2\, \vec{\epsilon} \cdot \vec{\epsilon}^{\,\,\prime}) \, ,
\end{equation}
where $\vec{k}$ is the momentum of the kaon in the $J/\psi \omega$  rest frame. Hence, the tree level amplitudes are  given by
\begin{eqnarray}\label{tree}
t^{tree}_{B^- \to K^- D^{*0} \bar{D}^{*0}} &=& \left[A\, |\vec{k}_{\rm av}|^2\, \vec{\epsilon}\cdot \vec{\epsilon}^{\,\,\prime}  \right.\, \nonumber\\
&+&  \left. B\, (\vec{\epsilon} \cdot \vec{k} \vec{\epsilon}^{\,\,\prime}\cdot \vec{k} -
\frac{1}{3} |\vec{k}|^2\, \vec{\epsilon} \cdot \vec{\epsilon}^{\,\,\prime}) \right] (1+3\,C) \, , \nonumber\\
t^{tree}_{B^- \to K^- D^{*+} {D}^{*-}} &=& A\, |\vec{k}_{\rm av}|^2\, \vec{\epsilon}\cdot \vec{\epsilon}^{\,\,\prime}  \, \nonumber\\
&+&  B\,  (\vec{\epsilon} \cdot \vec{k} \vec{\epsilon}^{\,\,\prime}\cdot \vec{k} -
\frac{1}{3} |\vec{k}|^2\, \vec{\epsilon} \cdot \vec{\epsilon}^{\,\,\prime})  \, , \nonumber\\
t^{tree}_{B^- \to K^- D_s^{*+} D_s^{*-}} &=& A\, |\vec{k}_{\rm av}|^2\, \vec{\epsilon}\cdot \vec{\epsilon}^{\,\,\prime} \, \nonumber\\
&+&  B\, (\vec{\epsilon} \cdot \vec{k} \vec{\epsilon}^{\,\,\prime}\cdot \vec{k} -
\frac{1}{3} |\vec{k}|^2\, \vec{\epsilon} \cdot \vec{\epsilon}^{\,\,\prime})    \, ,
\end{eqnarray}
where we  have substituted $A'$ of Eq.\eqref{A} by $A |\vec{k}_{\rm av}|^2$ with $\vec{k}_{\rm av}$, an average value of $\vec{k}$, just to make $A$ and  $B$ with the same dimension. We take $|\vec{k}_{\rm av}|=1000$ MeV.

Next, we take into account the rescattering depicted in Fig.~\ref{res2} to get $K^- J/\psi\omega$ in the final state  and we find the amplitude for $J/\psi\omega$  production
\begin{equation}\label{jomega}
t_{J/\psi\omega} = A\, |\vec{k}_{\rm av}|^2\, \vec{\epsilon}\cdot \vec{\epsilon}^{\,\,\prime}\,\, t_1 +
B (\vec{\epsilon} \cdot \vec{k} \,\,\vec{\epsilon}^{\,\,\prime}\cdot \vec{k} -
 \frac{1}{3} |\vec{k}|^2\, \vec{\epsilon} \cdot \vec{\epsilon}^{\,\,\prime})\,\, t_2 \, ,
\end{equation}
where $t_1$ and $t_2$ are given by
\begin{eqnarray}\label{t1}
t_1 &=& G_{D^{*0}\bar{D}^{*0}}(M_{\rm inv}) \, t^{I}_{D^{*0}\bar{D}^{*0}\to J/\psi\omega} (1+3\,C) \, \nonumber\\
 &+& G_{D^{*+} D^{*-}}(M_{\rm inv}) \, t^{I}_{D^{*+} D^{*-}\to J/\psi\omega}\, \nonumber\\
 &+& G_{D_s^{*+} D_s^{*-}}(M_{\rm inv}) \, t^{I}_{D_s^{*+} D_s^{*-}\to J/\psi\omega} \,,
\end{eqnarray}
\begin{eqnarray}\label{t2}
t_2 &=& G_{D^{*0}\bar{D}^{*0}}(M_{\rm inv}) \, t^{II}_{D^{*0}\bar{D}^{*0}\to J/\psi\omega} (1+3\,C) \, \nonumber\\
 &+& G_{D^{*+} D^{*-}}(M_{\rm inv}) \, t^{II}_{D^{*+} D^{*-}\to J/\psi\omega}\, \nonumber\\
 &+& G_{D_s^{*+} D_s^{*-}}(M_{\rm inv}) \, t^{II}_{D_s^{*+} D_s^{*-}\to J/\psi\omega} \,,
\end{eqnarray}
with $t^{I}$ the amplitude for the scalar resonance $J^{PC}=0^{++}$, $3940$ MeV and $t^{II}$
for the tensor resonance  $J^{PC}=2^{++}$,  $3930$ MeV.

 Since the $\vec{\epsilon}\cdot \vec{\epsilon}^{\,\,\prime}$ and $\vec{\epsilon}\cdot \vec{k}\,
\vec{\epsilon}^{\,\,\prime}\cdot \vec{k}-\frac{1}{3}|\vec{k}|^2\vec{\epsilon}\cdot \vec{\epsilon}^{\,\,\prime}$
structures filter spin $0$ and $2$ respectively, the structure is kept in the iterations
implicit in Eqs.~\eqref{t1} and \eqref{t2}. The $G$ functions in the former
equations are the vector-vector loop functions for the intermediate $D^*\bar{D}^*$, $D_s^*\bar{D}_s^*$
in Fig.~\ref{res2}. They are regularized in \cite{molina} using dimensional regularization
with the subtraction constant $a=-2.07$ and $\mu=1000$ MeV. Here, we follow the prescription
of \cite{Dai:2018tgo,Wang:2017mrt} and we use the cutoff method with ${q}_{max}$ fixed to reproduce the results of \cite{molina}.
 In the former equations $A$ and $B$
are functions (we take them as constants in the limited range of invariant mass studied) which have to do
with the weight of the weak process and hadronization before the final state interaction is
taken into account. We shall vary $A$ and $B$ within a reasonable range to see the results.

With the amplitudes of Eq.~\eqref{jomega} the mass distributions,
summing $|t|^2$ over the final vector polarizations, is given by
\begin{eqnarray}\label{dgdm}
\frac{d\Gamma}{dM_{\rm inv}^{J/\psi\omega}}& =& \frac{1}{(2\pi)^3}\frac{1}{4M_{B}^2} k^{\prime}
\tilde{p}_{\omega}  \nonumber\\
&\times & \Big( 3 |A|^2 |\vec{k}_{\rm av}|^4 |t_1|^2 +\frac{2}{3} |B|^2 |\vec{k}|^4 |t_2|^2\Big)  \, ,~~~
\end{eqnarray}
where $k^{\prime}$ is the kaon  momentum in the $B^-$ rest frame, $\tilde{p}_{\omega}$ the
$\omega$ momentum in the $J/\psi\omega$ rest frame and $k$ the kaon momentum in the
$J/\psi\omega$ rest frame for the $J/\psi\omega$ final state,
\begin{eqnarray}\label{ptil}
k^{\prime} = \frac{\lambda^{1/2}(M^2_{B^-},m^2_{K},M_{\rm inv}^{2})}{2M_{B^-}}\, , \nonumber\\
k = \frac{\lambda^{1/2}(M^2_{B^-},m^2_{K},M_{\rm inv}^{2})}{2M_{\rm inv}}\, , \nonumber\\
\tilde{p}_{\omega} = \frac{\lambda^{1/2}(M_{\rm inv}^{2},m^2_{J/\psi},m^2_{\omega})}{2 M_{\rm inv}}\, .
\end{eqnarray}

Next we need the  amplitudes $t^I$ and $t^{II}$ obtained from \cite{molina}, in which  the Flatt\'e form
of the amplitude was used in terms of the couplings and the width.  We have listed the couplings in Table \ref{tab:g1} and \ref{tab:g2}.

The amplitudes are,  $i=I$ or $II$,  given by
\begin{eqnarray}
t^{i}_{D^{*0}\bar{D}^{*0},\, J/\psi\omega} & =& \frac{\frac{1}{\sqrt{2}} g^{(i)}_{R,\,D^*\bar{D}^*}\,\, g^{(i)}_{R,\,J/\psi\omega}}{M^{2}_{\rm inv}
 - M^2_{R_i} + i M_{R_i}\Gamma_{R_i}} \, , \nonumber\\
 t^{i}_{D^{*+}D^{*-},\, J/\psi\omega} & =& \frac{\frac{1}{\sqrt{2}} g^{(i)}_{R,\,D^*\bar{D}^*}\,\, g^{(i)}_{R,\,J/\psi\omega}}{M^{2}_{\rm inv}
 - M^2_{R_i} + i M_{R_i}\Gamma_{R_i}} \, ,\nonumber\\
 t^{i}_{D_s^{*+}D_s^{*-},\, J/\psi\omega} & =& \frac{ g^{(i)}_{R,\,D_s^{*+}D_s^{*-}}\,\, g^{(i)}_{R,\,J/\psi\omega}}{M^{2}_{\rm inv}
 - M^2_{R_i} + i M_{R_i}\Gamma_{R_i}} \, ,
\end{eqnarray}
where the width $\Gamma_{R_i}$ is taken as
\begin{equation}\label{gamma}
\Gamma_{R_i} = \Gamma^{(i)}_0 + \Gamma^{(i)}_{J/\psi\omega} + \Gamma^{(i)}_{D^*\bar{D}^*}\, ,
\end{equation}
with
\begin{equation}
\Gamma_{J/\psi\omega}^{(i)} = \frac{|g^{i}_{R,\,J/\psi\omega}|^2}{8\pi M^2_{R_i}}\,\tilde{p}_{\omega}\, ,
\end{equation}
and $\tilde{p}_{\omega}$ given by Eq.~\eqref{ptil} as a function of $M_{\rm inv}$, and
\begin{equation}\label{gammadd}
\Gamma_{D^*\bar{D}^*}^{(i)} = \frac{|g^{i}_{R,\,D^*\bar{D}^*}|^2}{8\pi M^2_{R_i}}\,\tilde{p}_{D^*}
\Theta(M_{\rm inv} - 2 M_{D^*})\, ,
\end{equation}
with $\tilde{p}_{D^*}$ as $\tilde{p}_{\omega}$ in Eq.~\eqref{ptil} with the changes $M_{J/\psi}\to M_{D^*}$,
$M_{\omega}\to M_{\bar{D}^*}$. The width $\Gamma^{(i)}_0$ in Eq.~\eqref{gamma} accounts for the channels
different of $J/\psi\omega$ and $D^*\bar{D}^*$, mostly the light channels, such that $\Gamma^{(i)}_0$
is practically constant and we take
\begin{equation}
\Gamma^{(i)}_0 = \Gamma_{R_i} - \Gamma^{(i)}_{J/\psi\omega}(M_{\rm inv}^{J/\psi\omega}=M_{R_i})\, .
\end{equation}

Note that in Eq.~\eqref{gammadd}, $\Gamma^{(i)}_{D^*\bar{D}^*}$ only starts above the
$D^*\bar{D}^*$ threshold, but since the coupling of the resonance to this channel is so large, it
grows fast above threshold giving rise to the Flatt\'e effect.

\section{Results}
\label{sec:res}
\begin{figure}
	\begin{center}
		\includegraphics[width=0.54\textwidth]{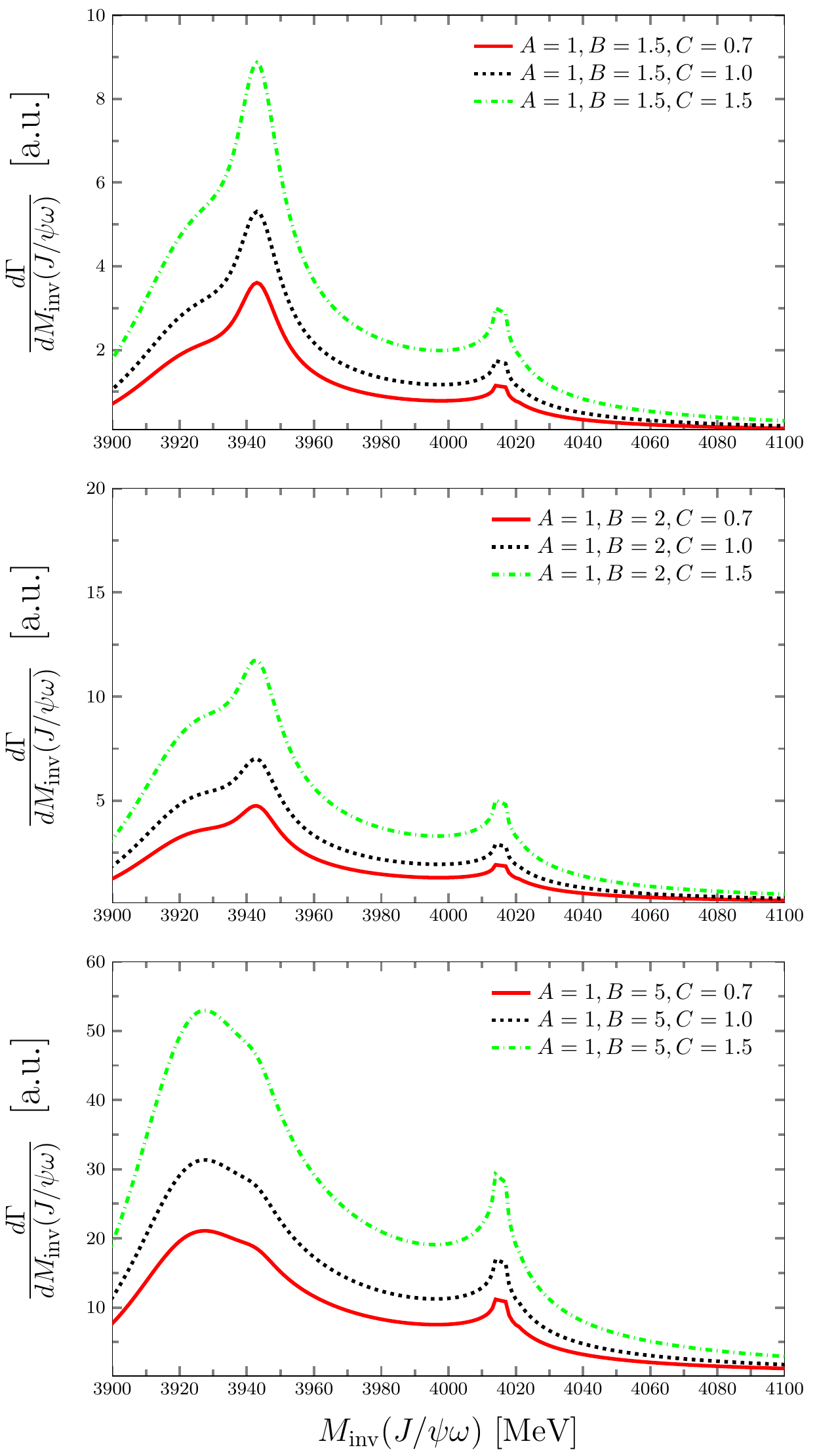}
	\end{center}
	\caption{\label{cern} $\frac{d\Gamma}{dM_{inv}^{J/\psi\omega}}$ the results for the different values of the
	parameter $B$ and $C$.}
\end{figure}

We will present the invariant mass distribution $\frac{d\Gamma}{dM_{inv}^{J/\psi\omega}}$ in arbitrary units.
Since $A$ and $B$ have been normalized to have the same dimensions, the two terms in Eq. \eqref{dgdm} have similar strength for $A=B$.
For this purpose, we take $A=1$ and look at the results for different values of $B$.
As discussed in \cite{Dai:2018tgo},  $\frac{B}{A} $ should be bigger than $1$, so we take the ratio of $1.5, 2, 5 $. As mentioned above,
we also  look at the results for different values of $C$.

 We show in Fig.\ref{cern} the results of $\frac{d\Gamma}{dM_{inv}^{J/\psi\omega}}$ for these different values. The absolute normalization
is arbitrary and the shape changes a bit since one gives more strength to one or another resonance changing $B$ and  $C$.
From Fig.\ref{cern} we can see that due to the proximity of the two resonances, and
the fact that both of them can be produced in this reaction, the two peaks actually
merge into a broader one, although a precise measurement could maybe allow a
separation of the two peaks, particularly if a partial wave analysis is done that separates
the two different spin resonances. Interesting, however, is the fact that the cusp appears
always at the same place, the $D^*\bar{D}^*$ threshold. The other relevant feature is that
its strength grows with increasing weight of the tensor resonance, indicating that the cusp
is basically tied to the $2^{++}$ $X(3930)$ state.

\begin{figure}
	\begin{center}
		\includegraphics[width=0.6\textwidth]{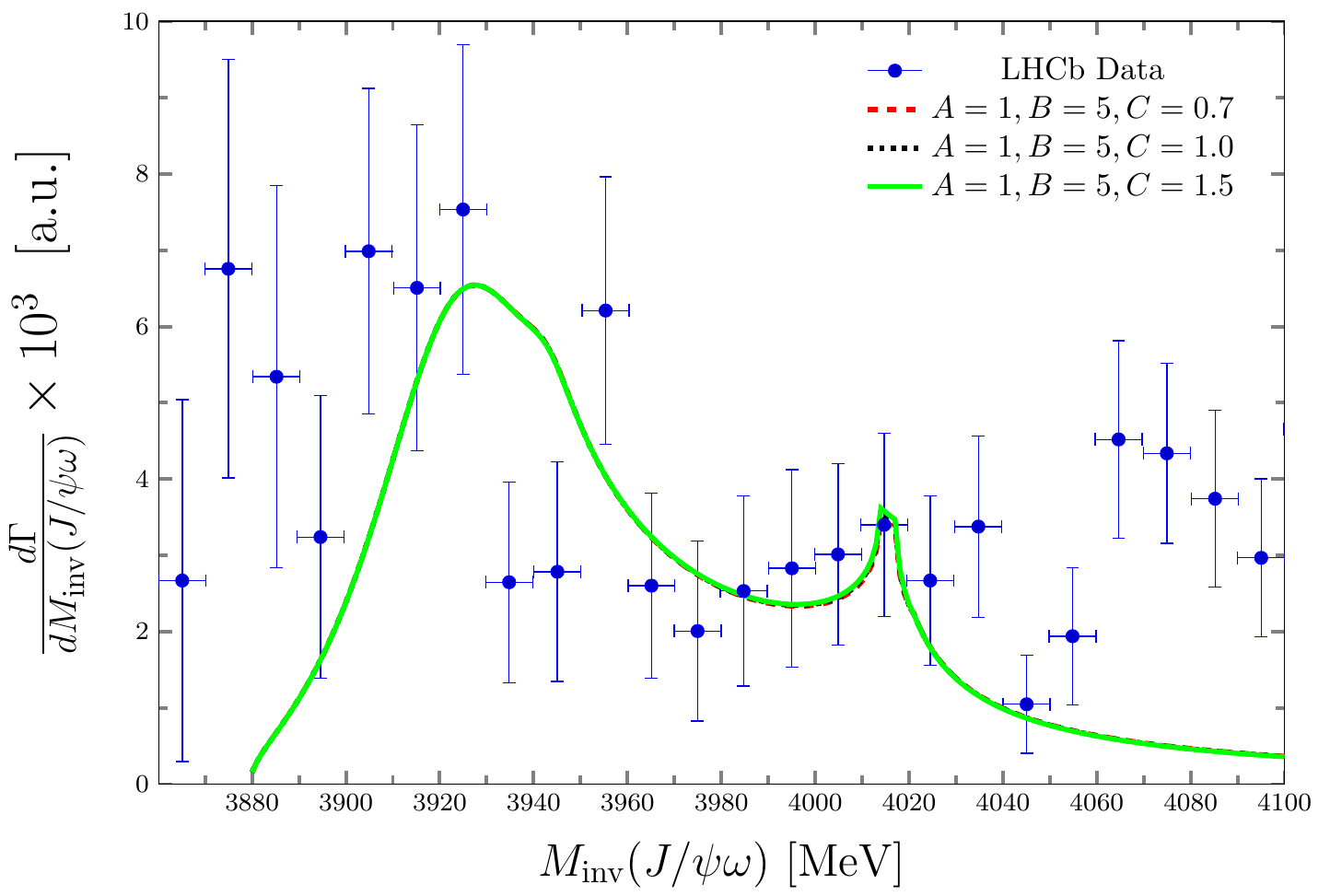}
	\end{center}
	\caption{\label{cerndata} Fitted $J/\psi \omega$ invariant mass distribution  to LHCb data \cite{Andreassi:2014skr} for $B^+ \to J/\psi \omega K^+ $ reaction  with different parameters  $B$ and $C$,  $A$ is normalized.
%for  $\frac{d\Gamma}{dM_{inv}^{J/\psi\omega}}$ decay
}
\end{figure}

In Fig. \ref{cerndata} we compare our results with the data of Ref. \cite{Andreassi:2014skr}.
%\ footnote{From the PHD thesis entitled "Search for exotic resonances in the decay $B^+ \to J/\psi \omega K^+ $ reaction in the LHCb experiment at CERN", we thank Dr. Guido Andreassi  for allowing us to use these data.}
 We have chosen $B=5$, which
gives a better reproduction of the data. The peak is due mostly to the tensor $X(3930)$ state and qualitatively accounts for the strength around $3900$ MeV observed in the experiment.
The other feature coming from  our framework is the unavoidable  cusp at the $D^* \bar{D}^*$ threshold. The data have large  errors but a fall down  of the distribution  beyond this threshold
is present in the data before it starts growing again, probably  due to the excitation  of other resonances that  are of a different  nature than the two discussed here.  One should note
the similarity of this feature  with the cusp at the $D_s^* \bar{D}_s^*$ threshold  observed in the  $B^+ \to J/\psi \omega K^+ $ experiment \cite{Aaij:2016iza}, which was interpreted
in \cite{Wang:2017mrt} in terms of resonances that couple to $D_s^* \bar{D}_s^*$ . The data also seems to indicate the contribution of another resonance around $3875$ MeV, which is
most probably the $X(3872)$. Once again, better  statistics and partial wave analyses will help shedding further light on this issue.

It could be most useful to get better data for this distribution, and the present work should serve as a motivation for it. Given the interpretation of the
 $B^+ \to J/\psi \omega K^+ $  data presented here, and its relevance to learn about the nature of some  resonances, it would be most advisable to increase the statistics
 in this reaction to improve the present precision.

\section{Conclusions}
\label{sec:conc}
We have presented a theoretical interpretation  of the data on the  $B^+\to J/\psi \omega K^+$ reaction~\cite{Andreassi:2014skr} in the range of $J/\psi \omega$ invariant mass $3900\sim 4050$ MeV. In this range we find two resonances $X(3930)$ and $X(3940)$   that couple strongly to $D^*\bar{D}^*$ in $J^{PC}=2^{++}, 0^{++}$ respectively.  The excitation of these resonances particularly the $X(3930)$  gives rise to a peak that accounts for
the large experimental strength around $3900$ MeV. The other feature is the unavoidable presence of a strong  cusp at the  $D^*\bar{D}^*$ threshold,
which seems to be supported  by experiment. This behaviour  is  reminiscent of the one observed in the  $B^+\to J/\psi \omega K^+$ reaction
where a cusp is observed at the  $D_s^* \bar{D}_s^*$ threshold, and which, similarly to the  present case, could also be interpreted in terms of resonances
coupled strongly to  $D_s^* \bar{D}_s^*$. The present discussion should serve to stimulate further experimental work, improving the statistics and disentangling the quantum numbers
of the structure observed.

\vskip .5cm
\section*{Acknowledgements}
LRD wishes to acknowledge the support from the National Natural Science Foundation of China
(No. 11575076) and the State Scholarship Fund of China (No. 201708210057).  This work is partly supported by
National Natural Science Foundation of China (No. 11505158).
This work is partly supported by the Spanish Ministerio
de Economia y Competitividad and European FEDER funds under Contracts No. FIS2017-84038-C2-1-P B
and No. FIS2017-84038-C2-2-P B, and the Generalitat Valenciana in the program Prometeo II-2014/068, and
the project Severo Ochoa of IFIC, SEV-2014-0398 (EO).

\end{document}